\documentclass[prl,aps,showpacs,epsf, twocolumn]{revtex4}

\def\be{\begin{equation}}
\def\ee{\end{equation}}

\usepackage{graphicx}
\usepackage{bm}
\usepackage{amsmath}
\usepackage{amssymb}
\usepackage{amsfonts}

\begin{document}

\title{Exploring the thermodynamics of a universal Fermi gas}

\author{S. Nascimb\`ene, N. Navon, K. J. Jiang, F. Chevy, and C. Salomon}

\affiliation{Laboratoire Kastler Brossel, CNRS, UPMC, \'Ecole
Normale Sup\'erieure, 24 rue Lhomond, 75231 Paris, France\\}


\maketitle

{\bf
From sand piles to electrons in metals, one of the greatest
challenges in modern physics is to understand the behavior of an
ensemble of strongly interacting particles.
A class of quantum many-body systems such as neutron matter and cold
Fermi gases share the same universal thermodynamic properties when
interactions reach the maximum effective value allowed by quantum
mechanics, the so-called unitary limit \cite{ho2004universal,inguscio2006ultracold}. It is
then possible to simulate some astrophysical phenomena inside the
highly controlled environment of an atomic physics laboratory.
Previous work on the thermodynamics of a two-component Fermi gas led
to thermodynamic quantities averaged over the trap
\cite{stewart2006potential,luo2007measurement,luo2009thermodynamic},
making it difficult to compare with many-body theories developed for
uniform gases.
Here we develop a general method that provides for the first time the equation of state of a uniform gas, as well as a detailed comparison with existing theories \cite{burovski2006critical,bulgac2006spin,haussmann2007thermodynamics,combescot09par,lobo2006nsp,liu2009virial,rupak2007universality,combescot2007nsh,combescot2008nsh,prokof'ev08fpb}. The precision of our equation of state leads to new physical insights on the unitary gas.
For the unpolarized gas, we show that the low-temperature
thermodynamics of the strongly interacting normal phase is well
described by Fermi liquid theory and we localize the superfluid transition. For a spin-polarized system \cite{shin2006observation,partridge2006pap,nascimbene2009pol}, our equation of state at zero temperature has a $2\%$ accuracy and it extends the work of \cite{shin2008pd,shin2008des} on the phase diagram to a new regime of precision.
We show in particular that, despite strong interactions, the normal phase behaves as a mixture of two ideal gases: a Fermi gas
of bare majority atoms and a non-interacting gas of dressed
quasi-particles, the fermionic polarons
\cite{shin2008des,chevy2006upa,lobo2006nsp,schirotzek2009ofp,nascimbene2009pol}.
}

In this letter we study the thermodynamics of a mixture of the two
lowest spin states ($i=1,2$) of $^6$Li prepared at a magnetic field
$B=834$ G (see Methods), where the dimensionless number $1/k_{F}a$
characterizing the $s$-wave interaction is equal to zero, the unitary limit. $k_F$ is
the Fermi momentum and $a$ the scattering length. Understanding the
universal thermodynamics at unitarity is a challenge for many-body
theories because of the strong interactions between particles. 
Despite this complexity at the microscopic scale, all the
macroscopic properties of an homogeneous system are encapsulated
within a single equation of state $P(\mu_1,\mu_2,T)$ that relates
the pressure $P$ of the gas to the chemical potentials $\mu_i$ of
the species $i$ and to the temperature $T$. In the unitary limit,
this relationship can be expressed as
\cite{ho2004universal}:
\begin{equation}\label{eqofstate}
P(\mu_1,\mu_2,T)=P_1(\mu_1,T)h\left(\eta=\frac{\mu_2}{\mu_1},\zeta=\exp\left(\frac{-\mu_1}{k_BT}\right)\right),
\end{equation}
where
$P_1(\mu_1,T)=-k_BT\lambda^{-3}_{dB}(T)f_{5/2}\left(-\zeta^{-1}\right)$
is the pressure of a single component non-interacting Fermi gas and
$f_{5/2}(z)=\sum_{n=1}^\infty z^n/n^{5/2}$. $h(\eta,\zeta)$ is a
universal function which contains all the thermodynamic information
of the unitary gas (Fig. 1). In cold atomic systems, the
inhomogeneity due to the trapping potential makes the measurement of
$h(\eta,\zeta)$ challenging. However, this inhomogeneity of the trap can be turned into an advantage as shown in \cite{shin2008des,ho2009opdtq}.

\begin{figure}
\vspace{-5mm}
\centerline{\includegraphics[width=0.8\columnwidth]{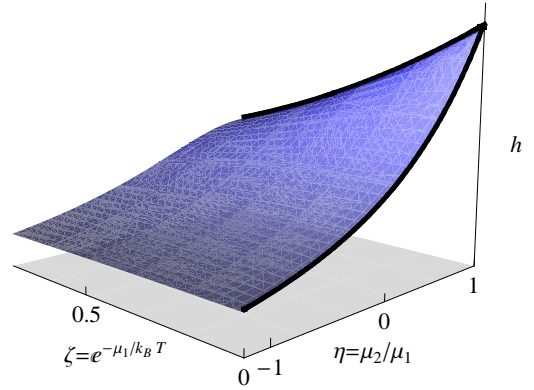}}
\vspace{-3mm} \caption{Schematic representation of the universal
function $h(\eta,\zeta)$. It fully describes the thermodynamics of
the unitary gas as a function of chemical potential imbalance
$\eta=\mu_2/\mu_1$ and of the inverse of the fugacity
$\zeta=\exp(-\mu_1/k_BT)$. In this paper we measure the function $h$
over the black lines $(1,\zeta)$ and $(\eta,0)$ which correspond to
the balanced unitary gas at finite temperature and to the
spin-imbalanced gas at zero temperature, respectively.} \label{Fig0}
\end{figure}

We directly probe the local pressure of the trapped gas using \emph{in situ} images, following the recent proposal \cite{ho2009opdtq}. In the local density
approximation, the gas is locally homogeneous with local chemical
potentials:
\begin{equation}
\mu_i(\mathbf{r})=\mu_i^0-V(\mathbf{r})\label{lda}
\end{equation}
($\mu_i^0$ is the chemical potential at the bottom of the trap for
species $i$). Then a simple formula relates the pressure $P$ to the
doubly-integrated density profiles \cite{ho2009opdtq}:
\begin{equation}
\label{Pn}
P(\mu_{1z},\mu_{2z},T)=\frac{m\omega_r^2}{2\pi}
\left(\overline{n}_1(z)+\overline{n}_2(z)\right),
\end{equation}
where $\overline{n}_i(z)=\int\;n_i(x,y,z)\textrm{d}x\textrm{d}y,$
$n_i$ being the atomic density. $\omega_r$ (resp. $\omega_z$) is the
transverse (resp. axial) angular frequency of a cylindrically
symmetric trap (see Fig. 2) and $\mu_{iz}=\mu_{i}(0,0,z)$ is
the local chemical potential along the $z$ axis.
From a single image, we thus measure the equation of state (\ref{eqofstate}) along the parametric line $(\eta,\zeta)=(\mu_{2z}/\mu_{1z},\exp(-\mu_{1z}/k_BT))$, see below.

\begin{figure}
\vspace{-3mm}
\centerline{\includegraphics[width=0.85\columnwidth]{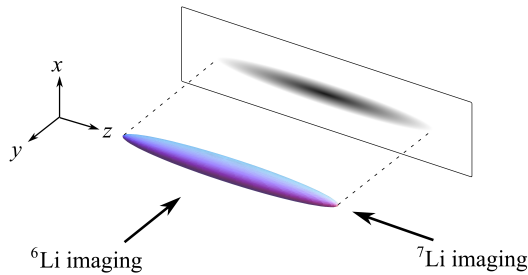}}
\vspace{-3mm} \caption{Schematic representation of our atomic sample. The $^6$Li atomic cloud is imaged in the
direction $y$; the column density is then integrated along the
direction $x$ to give $\overline{n}(z)$. The $^7$Li atoms are imaged
after a time of flight along the $z$ direction.}\label{Fig4}
\end{figure}

The interest of this method is straightforward. First, one directly
measures the equation of state (EOS) of the uniform gas. Second,
each pixel row $z_i$ gives a point $h(\eta(z_i),\zeta(z_i))$ whose
signal to noise ratio is essentially given by the one of
$\overline{n}_1(z)+\overline{n}_2(z)$; typically one experimental
run leads to $\sim100$ points with a signal to noise between 3 and
10. With about 40 images one gets $\simeq4000$ points
$h(\eta,\zeta)$, which after averaging provides a low-noise EOS of standard deviation $\sigma=2\%$.
In the following we illustrate the efficiency of our method on two
important sectors of the parameter space $(\eta,\zeta)$ in
Fig. 1: the balanced gas at finite temperature $(1,\zeta)$
and the zero-temperature imbalanced gas $(\eta,0)$.

We first measure the equation of state of the unpolarized unitary
gas at finite temperature, $P(\mu_1,\mu_2,T)=P(\mu,T)$. The
measurement of $h(1,\zeta)$ through the local pressure (\ref{Pn})
can be done provided one knows the temperature $T$ of the cloud and
its central chemical potential $\mu^0$.

In the balanced case, model-independent thermometry is
notoriously difficult because of the strong interactions. Inspired
by \cite{spiegelhalder2009collisional}, we overcome this issue by
measuring the temperature of a $^7$Li cloud in thermal equilibrium
with the $^6$Li mixture (see Methods).

$\mu^0$ is fitted on the hottest clouds so that the EOS
agrees in the classical regime $\zeta\gg1$ with the second-order
virial expansion $h(1,\zeta)\simeq2(1+\zeta^{-1}/\sqrt2)$
\cite{ho2004high}. For colder clouds we proceed recursively. The EOS
of an image recorded at temperature $T$ has some overlap with the
previously determined EOS from all images with $T'>T$. In this
overlap region $\mu^0$ is fitted to minimize the distance between
the two EOS's. This provides a new portion of the EOS at lower
temperature. Using 40 images of clouds prepared at different
temperatures, we thus reconstruct a low-noise EOS from the classical part down to the degenerate regime, as shown in Fig. 3a.

We now comment the main features of the equation of state. At high
temperature, the EOS can be expanded in powers of $\zeta^{-1}$ as a
virial expansion \cite{liu2009virial}:
$$
\frac{h(1,\zeta)}{2}=\frac{\sum_{k=1}^{\infty}\left((-1)^{k+1}k^{-5/2}+b_k\right)\zeta^{-k}}{\sum_{k=1}^{\infty}(-1)^{k+1}k^{-5/2}\zeta^{-k}},
$$
where $b_k$ is the $k^\mathrm{th}$ virial coefficient.
Since we have $b_2=1/\sqrt{2}$ in the measurement scheme described
above, our data provides for the first time the experimental values
of
$b_3$ and $b_4$.
$b_3=-0.35(2)$ is in excellent agreement with the recent calculation
$b_3=-0.291-3^{-5/2}=-0.355$ from \cite{liu2009virial} but not with
$b_3=1.05$ from \cite{rupak2007universality}. $b_4=0.096(15)$
involves the 4-fermion problem at unitarity and could interestingly
be computed along the lines of \cite{liu2009virial}.

Let us now focus on the low-temperature regime of the normal phase
$\zeta\ll1$. As shown in Fig. 3b, we observe a $T^2$ dependence of the pressure with temperature. This behavior is reminiscent of a Fermi liquid and indicates that pseudogap effects expected for strongly-interacting Fermi superfluids \cite{chen2005bcs} do not show up at the thermodynamic level within our experimental precision. In analogy with $^3$He or heavy-fermion metals, we fit our data with
the EOS:
\begin{equation}
P(\mu,T)=2P_1(\mu,0)\left(\xi_n^{-3/2}+\frac{5\pi^2}{8}\xi_n^{-1/2}\frac{m^*}{m}\left(\frac{k_BT}{\mu}\right)^{2}\right),\label{Sommerfeld}
\end{equation}
$P_1(\mu,0)=1/15\pi^2(2m/\hbar^2)^{3/2}\mu^{5/2}$ being the pressure
of a single-component Fermi gas at zero temperature. $m^*$ is the
quasi-particle mass and $\xi_n^{-1}$ is the
compressibility of the normal gas extrapolated to zero temperature, and normalized to that of an ideal gas of same density. We
deduce two new parameters $m^*/m=1.13(3)$ and $\xi_n=0.51(2)$.
Despite the strong interactions $m^*$ is close to $m$, unlike the weakly interacting
$^3$He liquid for which $2.7<m^*/m<5.8$, depending on pressure. Our $\xi_n$ value is in agreement with the
variational Fixed-Node Monte-Carlo calculations $\xi_n=0.54$ in \cite{carlson2003superfluid}, $\xi_n=0.56$ in
\cite{lobo2006nsp} and with the Quantum Monte-Carlo calculation $\xi_n=0.52$ in \cite{bulgac2008quantum}. This yields the Landau parameters $F_0^{s}=\xi_n m^*/m-1=-0.42$ and $F_1^{s}=3(m^*/m-1)=0.39$.

In the lowest temperature points (Fig. 3c) we observe a
sudden deviation of the data from the fit (\ref{Sommerfeld}) at
$(k_BT/\mu)_c=0.32(3)$ (see supplementary materials). We interpret this behavior as the
transition from the normal phase to the superfluid phase. This critical ratio has been extensively calculated in the recent years.
Our
value is in close agreement with the diagrammatic Monte-Carlo
calculation $(k_BT/\mu)_c=0.32(2)$ of \cite{burovski2006critical} and with the Quantum Monte-Carlo calculation $(k_BT/\mu)_c=0.35(3)$ of \cite{bulgac2008quantum}
but differs from the self-consistent approach in
\cite{haussmann2007thermodynamics} giving $(k_BT/\mu)_c=0.41$,
from the renormalization group prediction
$0.24$ in \cite{gubbels2008ren}, and from several other less precise
theories. From eq. (\ref{Sommerfeld}) we deduce the total
density $n=n_1+n_2=\partial P(\mu_i=\mu,T)/\partial\mu$ and the
Fermi energy $E_F=k_BT_F=\hbar^2/2m(3\pi^2n)^{2/3}$ at the
transition point. We obtain $(\mu/E_F)_c=0.49(2)$ and
$(T/T_F)_c$=0.157(15), in very good agreement with
\cite{burovski2006critical}. Our measurement is the first direct
determination of $(\mu/E_F)_c$ and $(T/T_F)_c$ in the homogeneous
gas. It agrees with the extrapolated value of the MIT measurement
\cite{shin2008pd}.

Below $T_c$, advanced theories
\cite{bulgac2006spin,haussmann2007thermodynamics} predict
that $P(\mu,T)/2P_1(\mu,0)$ is nearly constant (Fig. 3b).
Therefore at $T=T_c$, $P/2P_1\simeq \xi_s^{-3/2}\simeq3.7$, and is
consistent with our data. Here $\xi_s=0.42(1)$ is the fundamental
parameter characterizing the EOS of the balanced superfluid at zero
temperature, a quantity extensively measured and computed in the
recent years \cite{inguscio2006ultracold}.

Our data is compared at all temperatures with the calculations from
\cite{burovski2006critical,bulgac2006spin,haussmann2007thermodynamics,combescot09par}
(Fig.3a). The agreement with
\cite{bulgac2006spin} is very good for a large range of
temperatures. Concerning \cite{burovski2006critical}, the deviation with our
data is about one error bar of the Monte-Carlo method below
$\zeta=0.2$ and the deviation increases with temperature
(Fig. 3a). Furthermore, we show in the supplementary
material that $h(1,\zeta)/2$ must be greater than 1, an inequality
violated by the two hottest Monte-Carlo points of
\cite{burovski2006critical}.

From our homogeneous EOS we can deduce the equation of state of the
harmonically trapped unitary gas by integrating $h(1,\zeta)$ over
the trap (see supplementary material). In particular, we find a
critical temperature for the trapped gas $(T/T_F)_c=0.19(2)$, where
$T_F=\hbar(3\omega_r^2\omega_zN)^{1/3}$. This value agrees very well
with the recent measurement of \cite{riedl2009superfluid}, and with
less precise measurements
\cite{greiner2003emergence,luo2009thermodynamic,inada2008critical}.
\begin{figure}
\vspace{-2mm}
\centerline{\includegraphics[width=\columnwidth]{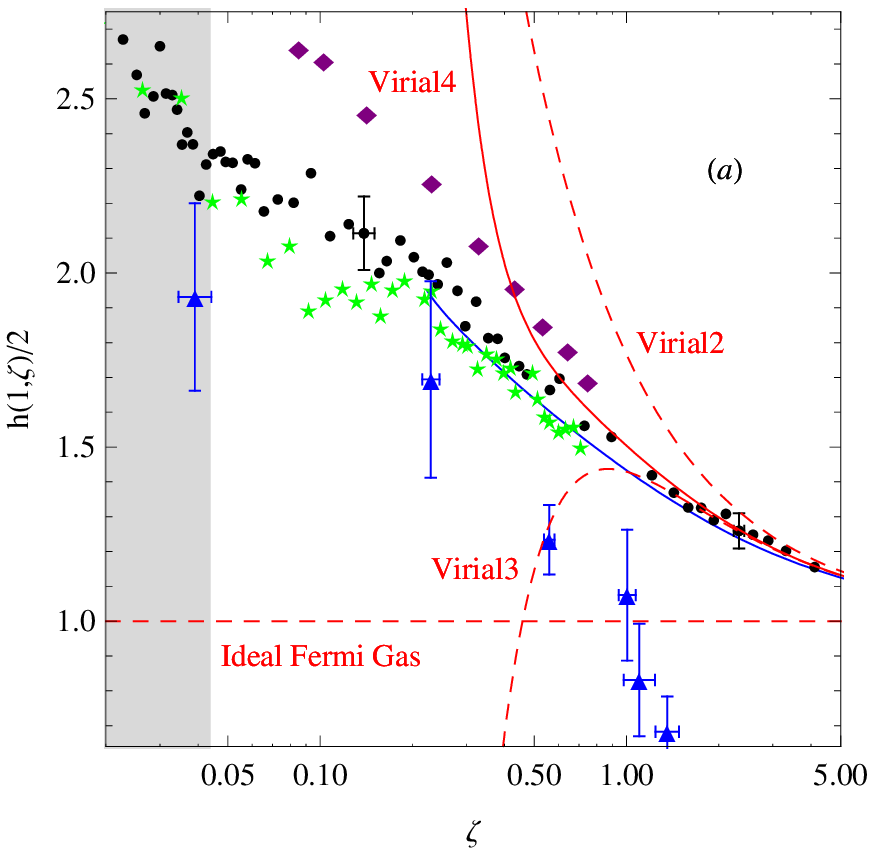}}
\vspace{-2mm}
\centerline{\includegraphics[width=\columnwidth]{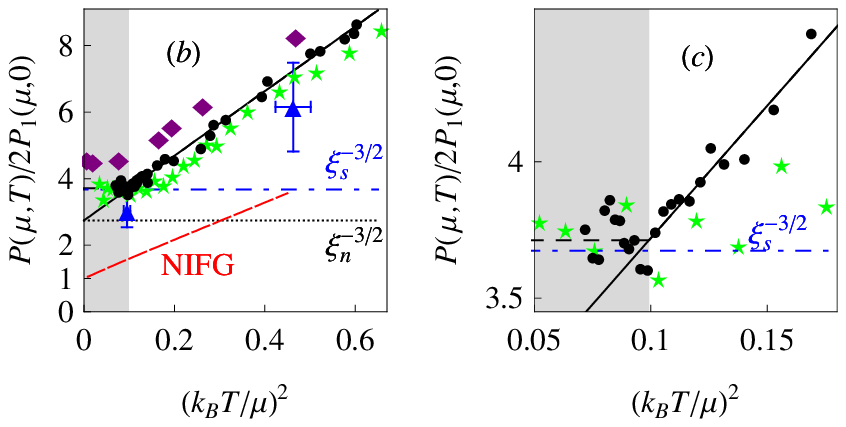}}
\vspace{-3mm} \caption{(color online) Equation of state of a spin-balanced unitary Fermi gas. (a) Finite-temperature
equation of state $h(1,\zeta)$ (black dots). The error bars represented at $\zeta=0.14$ and
$\zeta=2.3$ indicate the $6\%$ accuracy in $\zeta$ and $h$ of our EOS. The red curves
are the successive virial expansions up to $4^{\mathrm{th}}$ order.
The blue triangles are from \cite{burovski2006critical}, the green
stars from \cite{bulgac2006spin}, the purple diamonds from
\cite{haussmann2007thermodynamics}, and the blue solid line from \cite{combescot09par}. The grey region indicates the
superfluid phase. (b) Equation of state $P(\mu,T)/2P_1(\mu,0)$ as a
function of $(k_BT/\mu)^2$, fitted by the Fermi liquid equation of
state (\ref{Sommerfeld}). The red dashed line is the non-interacting
Fermi gas (NIFG). The horizontal dot-dashed (resp. dotted) line
indicates the zero-temperature pressure of the superfluid phase
$\propto\xi_s^{-3/2}$ (resp. normal phase $\propto\xi_n^{-3/2}$).
(c) Expanded view of (b) near $T_c$. The sudden deviation of the
data from the fit occurs at $(k_BT/\mu)_c=0.32(3)$ that we interpret
as the superfluid transition. The black dashed line indicates the mean value of the data points below $T_c$.} \label{Fig1}
\end{figure}

Let us now explore a second line in the universal diagram
$h(\eta,\zeta)$ (Fig. 1) by considering the case of the
$T=0$ spin-imbalanced mixture $\mu_2\neq\mu_1$, \emph{i.e.}
$\eta\neq1$. Previous work
\cite{shin2006observation,partridge2006pap,nascimbene2009pol} has
shown that phase separation occurs in a trap. Below a critical
population imbalance a fully-paired superfluid occupies the center
of the trap. It is surrounded by a normal mixed phase and an outer
rim consisting of an ideal gas of the majority component. In two out
of the three previous experiments including ours
\cite{shin2006observation,nascimbene2009pol}, the local density
approximation has been carefully checked. We are therefore entitled
to use (\ref{Pn}) to analyze our data.

As in the previous case, the relationship between the pressure and
the EOS requires the knowledge of the chemical potentials $\mu_1^0$
and $\mu_2^0$ at the center of the trap.

$\mu_1^0$ is determined using the outer shell of the
majority spin component ($i=1$). The pressure profile
$P(\mu_{1z},\mu_{2z},0)$ corresponds to the Fermi-Dirac distribution
and is fitted with the Thomas-Fermi formula
$P_1=\alpha(1-z^2/R_1^2)^{5/2}$, providing
$\mu_1^0=\frac{1}{2}m\omega_z^2R_1^2$. Using $P_1$ for the
calculation of $h=P/P_1$ cancels many systematic effects on the
absolute value of the pressure. Moreover, fitting the outer shell using a finite-temperature Thomas-Fermi profile \cite{shin2008pd}, we measure a temperature $k_BT=0.03(3)\mu_1^0$.

$\mu_2^0$ is fitted by comparison in the superfluid
region with the superfluid equation of state at zero temperature
\cite{chevy2006upa}:
\begin{equation}
h(\eta,0)=(1+\eta)^{5/2}/(2\xi_s)^{3/2}.\label{eqsuperfluid}
\end{equation}

\begin{figure}
  \vspace{-1mm}
\centerline{\includegraphics[width=\columnwidth]{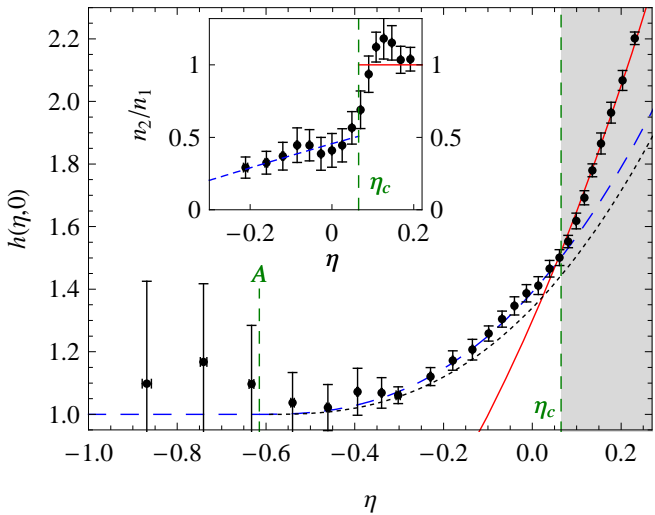}}
\vspace{-3mm}
\caption{(color online) Equation of state of the zero-temperature spin-imbalanced unitary gas $h(\eta,0)$ (black dots). Error bars are equal to one standard error. The red solid line is the
superfluid equation of state, the blue dashed line is the ideal
Fermi liquid equation (\ref{pressure_normal}) with $A=-0.615$,
$m^*=1.20m$ and the black dotted line is the Monte Carlo calculation
from \cite{lobo2006nsp}. Inset: Local density ratio $n_2/n_1$ as a
function of $\eta$. The red solid line $n_2/n_1=1$ corresponds to the
fully paired superfluid and blue dashed line to the model
(\ref{pressure_normal}).}\label{Fig2}
\end{figure}

Our measured equation of state $h(\eta,0)$ is displayed in
Fig.4. By construction our data agrees for $\eta\gtrsim0.1$ with eq.(\ref{eqsuperfluid}). In Fig.4 the slope of $h(\eta,0)$ displays an obvious discontinuity for
$\eta=\eta_c=0.065(20)$. This is a signature of a first-order
quantum phase transition to the partially polarized normal phase.
The error bar is dominated by the uncertainty on $\xi_s$. This value
is slightly higher than the prediction $\eta_c=0.02$ given by the
fixed-node Monte-Carlo \cite{lobo2006nsp} and than the value
$\eta_c=0.03(2)$ measured in \cite{shin2008pd}.

From the relations $n_i=\partial P/\partial\mu_i$ we deduce from
$h(\eta,0)$ the density ratio $n_2/n_1$ (see inset in
Fig.4). This ratio is discontinuous at the phase transition, from a maximum
value in the normal phase $(n_2/n_1)_c=0.5(1)$ to $n_2=n_1$ in the
superfluid phase. Our value is close to the
zero-temperature calculation 0.44 \cite{lobo2006nsp} and agrees with
the coldest MIT samples \cite{shin2008pd,shin2008des}. It confirms that the temperature is much smaller than the tricritical point temperature $T=0.07T_F$ \cite{shin2008pd} where the discontinuity vanishes, justifying
our $T=0$ assumption made above.

For $\eta<\eta_c$ our data displays a good agreement with a simple polaron model, based on the pioneering work in \cite{lobo2006nsp}. A
polaron is a quasi-particle describing a single minority atom
immersed in the majority Fermi sea
\cite{chevy2006upa,prokof'ev08fpb,schirotzek2009ofp,nascimbene2009pol}. It is characterized by a
renormalized chemical potential $\mu_2-A\,\mu_1$ and an effective
mass $m^*_p$ \cite{lobo2006nsp}. Following this picture, we write the pressure as the sum of the Fermi pressures of ideal gases of majority atoms and of polarons: 
\begin{equation}\label{Ppol}
P=\frac{1}{15\pi^2}\left(\frac{2m}{\hbar^2}\right)^{3/2}\left(\mu_1^{5/2}+\left(\frac{m_p^*}{m}\right)^{3/2}(\mu_2-A\mu_1)^{5/2}\right),
\end{equation}
which can be written as
\begin{equation}
h(\eta,0)=1+\left(\frac{m^*_p}{m}\right)^{3/2}(\eta-A)^{5/2}.\label{pressure_normal}
\end{equation}
$A$ and $m^*_p$ have recently been calculated exactly
\cite{prokof'ev08fpb,combescot2008nsh}: $A=-0.615$, $m^*_p/m=1.20(2)$ and with
these values inserted in (\ref{pressure_normal}) the agreement with
our data is perfect. Note that our data lies slightly
above the variational fixed-node Monte Carlo calculation
\cite{lobo2006nsp}.
We therefore conclude that interactions between polarons are not visible at this level of precision.

Alternatively, we can fit our data with $m^*_p/m$ as a free
parameter in (\ref{pressure_normal}). We obtain $m^*_p/m=1.20(2)$.
The uncertainty combines the standard error of the fit and the
uncertainty on $\xi_s$. This value agrees with our previous
measurement $m^*_p/m=1.17(10)$ \cite{nascimbene2009pol} (with a
5-fold improvement in precision), with the theoretical value
$m^*_p/m=1.20(2)$ in \cite{prokof'ev08fpb,combescot2008nsh} and with the
variational calculation \cite{combescot2007nsh}. It differs from the
values $1.09(2)$ in \cite{pilati2008psi}, $1.04(3)$ in
\cite{lobo2006nsp}, and from the
experimental value $1.06$ in \cite{shin2008des}.

We arrive at a simple physical picture of the $T=0$ spin-polarized
gas: the fully paired superfluid is described by an ideal gas EOS
renormalized by a single coefficient $\xi_s$; the normal phase is
nothing but two ideal gases, one of bare majority particles and one
of polaronic quasi-particles.

\begin{table}
  \centering
  \vspace{-1mm}
  \begin{tabular}{ccccc}
              \hline
              \hline
              $b_3$   & $b_4$     & $(k_BT/\mu)_c$ & $(\mu/E_F)_c$ & $(T/T_F)_c$   \\
              -0.35(2)& 0.096(15)  & 0.32(3)        & 0.49(2)       & 0.157(15)      \\
              \hline
              $\xi_n$ & $m^*/m$& $\eta_c$ & $(n_2/n_1)_c$&$m^*_p/m$ \\
              0.51(2) & 1.13(3)& 0.065(20) & 0.5(1)&1.20(2) \\
              \hline
              \hline
  \end{tabular}
  \caption{Table of quantities measured in this work.}\label{tableau}
\end{table}

In conclusion, we have introduced a powerful method for the
measurement of the equation of state of the unitary and homogeneous
Fermi gas, that enables direct comparison with theoretical models
and provides a set of new parameters shown in Tab.\ref{tableau}.
 The method can readily be extended to any multi-component cold atom gas in three dimensions that fulfills the local
density approximation (see supplementary discussion).
We have shown that the normal phase of the unitary Fermi gas is a strongly correlated system whose thermodynamic properties are well described by Fermi liquid theory, unlike
high-$T_c$ cuprates.

\emph{Note added in proof}: Since this paper was accepted for publication, we
have become aware of the measurement of a similar equation of state
for the balanced unitary Fermi gas at finite temperature by different
methods \cite{horikoshi2010measurement}.

\smallskip
\noindent\textbf{METHODS SUMMARY}\\
Our experimental setup is presented in \cite{nascimbene2009pol}. We
load into an optical dipole trap  and evaporate a mixture of $^6$Li in the
$|1/2, \pm1/2\rangle$ states and of $^7$Li in the $|1,1\rangle$
state at 834~G. The cloud typically contains
$N_6=5$~to~$10\times10^4$ $^6$Li atoms in each spin state and
$N_7=3$~to~$20\times10^3$ $^7$Li atoms at a temperature from $T=150$~nK to 1.3~$\mu$K. The $^6$Li trap frequencies
are $\omega_z/2\pi=37$~Hz, $\omega_r/2\pi$ varying from 830~Hz to 2.20~kHz, and the trap depth is 25~$\mu$K for our hottest samples, with $T\simeq2T_F$. $^6$Li atoms are imaged \emph{in situ} using
absorption imaging, while $^7$Li atoms are imaged after time of
flight, providing the temperature in the same experimental run (Fig. 4).
Since the scattering length describing the interaction between
$^7$Li and $^6$Li atoms, $a_{67}=2$~nm, is much smaller than
$k_F^{-1}$, the $^7$Li thermometer has no influence on the $^6$Li
density profiles. The $^7$Li-$^6$Li collision rate,
$\Gamma_{67}=10$~s$^{-1}$, is large enough to ensure thermal
equilibrium between the two species. 

\bibliographystyle{naturemag}



\bigskip
\noindent\textbf{Acknowledgements} We are grateful to R. Combescot, X. Leyronas, Y. Castin, A. Recati, S. Stringari, S. Giorgini, M. Zwierlein and T. Giamarchi for fruitful discussions and to C. Cohen-Tannoudji, J. Dalibard, F. Gerbier and G. Shlyapnikov for
critical reading of the manuscript. 
We acknowledge support from ESF (Euroquam),
SCALA, ANR FABIOLA, R\'egion Ile de France (IFRAF), ERC and Institut
Universitaire de France.

\bigskip
\noindent\textbf{Author Contributions} S. Nascimb\`ene and N. Navon
contributed equally to this work. S.N., N.N. and K.J. took the experimental data and all authors contributed to the data analysis and writing of the manuscript.

\bigskip
\noindent\textbf{Author Information} Correspondence and requests for
materials should be addressed to S. N. ~(email:sylvain.nascimbene@ens.fr).

\bigskip
\noindent\textbf{METHODS}\\
\noindent\textbf{Construction of the equation of state by successive patches.} A typical image at high temperature provides about 100 pixels corresponding to $\zeta$ values varying from 2 at the trap center to 6 at the edges, with a signal-to-noise from 3 to 10. 7 such images are fitted in the wings using the second-order virial expansion and	averaged to obtain a low-noise EOS up to $\zeta=2$. Then images of clouds where the evaporation has been pushed to a slightly lower temperature are recorded. They show about 75$\%$ overlap in $\zeta$ with the previous EOS. After minimization of the distance between a new image and the previously determined EOS in the overlap region, we obtain the value of $\mu^0$ for a single image with 3$\%$ statistical uncertainty. This process is repeated for 6 successive trap depths. When averaging one image with typically 10 previous images, we obtain a new EOS with an error on $\zeta$ of about $0.03/\sqrt{10}\simeq1\%$. The EOS experiences a random walk error on the 40 images of $0.01\times\sqrt{40}\simeq5\%$ for the coldest data. An independent check of the maximum error is provided by the good agreement with the superfluid equation of state for temperatures lower than $T_c$ \cite{bulgac2006spin,haussmann2007thermodynamics}.\\

\smallskip
\noindent\textbf{Evaluation of the systematic uncertainties.} For the measurement of $h(1,\zeta)$, the combined uncertainties on the radial frequency of the trap,  trap anharmonicity,  magnification of our imaging system, and atom counting affect the pressure measurement given in (\ref{Pn}) at $\simeq20\%$ level.
However, two measurements, one at relatively high temperature and one at very low temperature, enable us to show that the overall error does not exceed 6$\%$. In the temperature range $\zeta>0.5$, the agreement
between the experimental value $b_3=-0.35(2)$ and the theoretical
value $b_3=-0.355$ of the third virial coefficient indicates that the global
systematic error is smaller than $6\%$. 
Second, at very low temperature, theory \cite{bulgac2006spin,haussmann2007thermodynamics} predicts that the variation of $P/2P_1$ as a function of $k_BT/\mu$ in the superfluid phase remains smaller than 5$\%$. Our value of $P/2P_1=3.75$ below the critical point is within 5$\%$ of the $T=0$ prediction $\xi_s^{-3/2}=3.7(2)$. This confirms that systematic errors for our coldest samples are also smaller than $6\%$. 

For the determination of the critical transition to superfluidity we fit the low-temperature data $P(\mu,T)/2P_1(\mu,0)$ with a variable horizontal line for $T<T_c$ and with the Fermi-liquid equation (\ref{Sommerfeld}) for $T>T_c$. The result of the fit is the dashed black line in Fig. 3c, which intersects equation (\ref{Sommerfeld}) at $(k_BT/\mu)_c=0.315(8)$. This statistical error is negligible compared to the error induced by the 6$\%$ systematic uncertainty discussed above, justifying our very simplified fit procedure. Indeed a 6$\%$ error on the pressure induces a 10$\%$ error on $\mu$ for images recorded in the vicinity of the critical temperature, leading to $(k_BT/\mu)_c=0.32(3)$. 

For the measurement of $h(\eta,0)$, the fit of the fully polarized wings of the cloud serves as a pressure calibration for the rest of the cloud, cancelling many systematic effects. 

In order to estimate temperature effects in the polarized gas, let us first remark that in the superfluid phase corrections scale as $T^4$ for the bosonic excitations and are exponentially suppressed by the gap for the fermionic ones \cite{bulgac2006spin}. So in our temperature range $k_BT=0.03\mu_1^0$ their contributions will be very small. On the other hand, in the partially polarized normal phase, we expect a typical Fermi liquid $T^2$ scaling. In order to obtain an estimate of the error on the EOS, we develop the following simple model. In equation (\ref{Ppol}) which describes a mixture of zero-temperature ideal gases, we replace the Fermi pressures by the finite-temperature pressures of ideal gases (see equation (\ref{eqofstate})):
\[
P(\mu_1,\mu_2,T)=P_1(\mu_1,T)+\left(\frac{m_p^*}{m}\right)^{3/2}P_1(\mu_2-A\mu_1,T),
\]
and run the analysis described in the main text. At $T=0.05\mu_1^0$, the correction on $h$ is less than $1\%$, half of our current error bar.

\smallskip
\noindent\textbf{Limit of $^7$Li Thermometry.} As the scattering length
between the $^7$Li atoms, $a_{77}=-3$~nm is negative, the $^7$Li cloud
becomes unstable when a BEC forms. This occurs at $T\sim150$~nK with typically 3500 atoms. Precise thermometry with lower atom numbers becomes difficult. For the measurement of the
zero-temperature equation of state of the imbalanced gas, we do not
use $^7$Li thermometry but rather the fit of the wings of the majority spin component.

\newpage

\section{Supplementary Discussion}

\begin{figure}[h!]
\centerline{\includegraphics[width=\columnwidth]{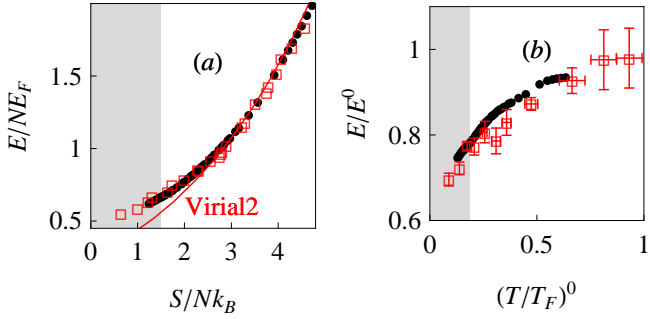}}
\caption{(color online) Equations of state of the trapped unitary gas. (a) Comparison between our EOS
$E/NE_F$ as a function of $S/Nk_B$ (black dots) and the EOS measured in \cite{luo2009thermodynamic} (open red squares).
The red solid line is the second-order virial equation of state. (b)
Comparison between our EOS $E/E^0$ as a function of $(T/T_F)^0$ (black dots) and the EOS measured on $^{40}$K in
\cite{stewart2006potential} (open red squares). The grey regions correspond to
the superfluid phase.}\label{Fig5}
\end{figure}


\subsection{Equation of State of the Trapped Unitary Gas}
In this work, we have measured the equation of state of the
homogeneous unitary gas. We can deduce from our data the EOS of the
trapped balanced unitary gas, which has been measured in
\cite{stewart2006potential,luo2009thermodynamic}.

Using the local density approximation, the total atom number $N=\int
n\,\textrm{d}r^3$ is expressed as a function of the temperature $T$
and the chemical potential $\mu^0$ at the center, involving the
function $h(1,\zeta)$:
\begin{equation}\label{Ntrapped}
N=\frac{-2}{\sqrt\pi}\left(\frac{k_BT}{\hbar\omega}\right)^{3}\int_{\zeta_0}^\infty\frac{\textrm{d}\log^{1/2}(\zeta/\zeta_0)}{\textrm{d}\zeta}f_{5/2}(-\zeta^{-1})h(\zeta)\textrm{d}\zeta,
\end{equation}
where $\zeta_0=\exp(-\mu^0/k_BT)$ and
$\omega=(\omega_r^2\omega_z)^{1/3}$. We use for the calculation a
discretized version of (\ref{Ntrapped}) taken solely on our
experimental values of $h$, \emph{i.e.} without using any
interpolating or fitting function. Similar expressions are used to calculate the Fermi temperature
$E_F=k_BT_F=\hbar\omega(3N)^{1/3}$, the total entropy $S$ and energy
$E$ of the cloud. The equation of state $E/NE_F$ as a function of
$S/Nk_B$, displayed in Supplementary Fig.\ref{Fig5}a, is in very good agreement
with \cite{luo2009thermodynamic}.

The normal-superfluid phase transition for the trapped gas occurs
when at the trap center $\zeta_0=\zeta_c=\exp(-(k_BT/\mu)_c^{-1})$,
with $(k_BT/\mu)_c=0.32(3)$, as measured on the homogeneous EOS
$h(1,\zeta)$. At this point we get $(T/T_F)_c=0.19(2)$,
$(S/Nk_B)_c=1.5(1)$ and $(E/NE_F)_c=0.67(5)$.

In order to make the comparison with \cite{stewart2006potential}, we
also express the equation of state $E/E^0$ as a function of
$(T/T_F)^0$, where the superscript $^0$ refers to the quantities
evaluated on a non-interacting Fermi gas having the same entropy
(Supplementary Fig.\ref{Fig5}b). The good agreement with the measurement in
\cite{stewart2006potential}, performed on $^{40}$K clouds,
illustrates the universality of the unitary gas. \\

\subsection{Physical interpretation of the pressure in the normal mixed phase}

We have shown that the pressure in the normal mixed phase can be described as the sum of the Fermi pressures of ideal gases of majority atoms and of polarons:
\begin{equation}\label{Ppol}
P=\frac{1}{15\pi^2}\left(\frac{2m}{\hbar^2}\right)^{3/2}\left(\mu_1^{5/2}+\left(\frac{m_p^*}{m}\right)^{3/2}(\mu_2-A\mu_1)^{5/2}\right).
\end{equation}
Here, we evaluate the corresponding canonical EOS relating the energy $E$ to the densities $n_1,n_2$, in order to compare with the Fixed Node Monte Carlo prediction \cite{lobo2006nsp}. Since at unitarity we have $E=3 PV/2$ \cite{ho2004universal}, we just  express the chemical potentials in terms of densities by using the thermodynamical identities $n_i=\partial_{\mu_i}P$, which yield respectively:
\begin{eqnarray}
n_2&=&\frac{1}{6\pi^2}\left[\frac{2m^*}{\hbar^2}\left(\mu_2-A\mu_1\right)\right]^{3/2}\\
n_1&=& \frac{1}{6\pi^2}\left(\frac{2m\mu_1}{\hbar^2}\right)^{3/2}-An_2. \label{n1}
\end{eqnarray}
The last term in equation (\ref{n1}) clearly indicates the increased majority density due to the presence of the minority component.
Expressing the pressure as a function of  $n_i$ in  (\ref{Ppol})  yields the energy:
$$E=E_{\rm FP}\left[\left(1+Ax\right)^{5/3}+\frac{m}{m^*}x^{5/3}\right],$$
\noindent where $E_{FP}$ is the energy of the fully polarized gas and $x=n_2/n_1$. Expanding $E$ to order $x^2$ finally leads to an expression similar to that obtained in \cite{lobo2006nsp}:
$$E(x)=E_{\rm FP}\left(1+\frac{5}{3}Ax+\frac{m}{m^*}x^{5/3}+Bx^2+....\right),$$
with $B=5A^2/9=0.2$. Our value of $B$ is close to the calculated value $B\simeq 0.14$ from \cite{lobo2006nsp}.\\

\subsection{Trap Anharmonicity}

First, in the axial direction $z$, the confinement is produced
magnetically and the corresponding anharmonicity is negligible. In
the radial direction, we develop the gaussian potential to fourth
order around $\rho=0$:
$$
V_r(\rho)=V_0\left(1-\exp\frac{-\rho^2}{\sigma^2}\right)\simeq\frac{1}{2}m\omega_r^2\rho^2+\epsilon\rho^4,
$$
where $m\omega_r^2=2V_0/\sigma^2$ and $\epsilon=-V_0/2\sigma^4$. In
the balanced case, we have
$$
\overline{n}(z)=\int\textrm{d}^2\rho\,
n\left(\mu^0-\frac{1}{2}m\omega_z^2z^2-\frac{1}{2}m\omega_r^2\rho^2-\epsilon\rho^4\right).
$$
Introducing $n=\partial P/\partial\mu$ and defining
$u=m\omega_r^2\rho^2/2+\epsilon\rho^4$ we obtain, to lowest order,
$$
\frac{m\omega_r^2}{2\pi}\overline{n}(z)=P(\mu_z)+\int_0^\infty
P(\mu_z-u)\frac{\textrm{d}u}{V_0}.
$$
The error on the measurement of $h$ is then
\begin{equation}\label{anharmonicites}
\frac{m\omega_r^2\,\overline{n}(z)}{2\pi
P_1(\mu_z,T)}-h(1,\zeta)=\frac{k_BT}{V_0}\int_\zeta^\infty\frac{f_{5/2}(-\zeta'^{-1})}{f_{5/2}(-\zeta^{-1})}\frac{h(1,\zeta')}{\zeta'}\textrm{d}\zeta'.
\end{equation}
We evaluate the integral in (\ref{anharmonicites}) using the
experimental values of $h(1,\zeta)$. In our shallowest trap, the
worst case anharmonicity effect is $5\%$. \\

\subsection{An exact inequality on the equation of state of an attractive Fermi gas}

Writing the hamiltonian as $\widehat H=\widehat H_0+\widehat U$,
where $\widehat H_0$ is the single-particle part of the hamiltonian
and $\widehat U$ is the inter-particle interaction, one has the
general inequality $ \Omega\le \Omega_0+\langle V\rangle_0$, where
$\Omega_0$ is the grand potential associated with $\widehat H_0$ and
$\langle\cdot\rangle_0$ is the thermal average related to $\widehat
H_0$
 \cite{feynman1972statistical}. Taking for $U$ a short range square potential of depth $U_0<0$
recovering the true scattering length, one has trivially
$\langle\widehat V\rangle_0<0$, hence $\Omega\le \Omega_0$. Using
the thermodynamic identity $\Omega=-PV$, and recalling that
$\Omega_0=-2P_1V$ and $h=P/P_1$, we finally get the inequality
$$h(1,\zeta)\ge2.$$ \\

\subsection{Extension to a Multi-Component System}

We extend the equation $(2)$ to a mixture of species $i$, of mass
$m_i$, trapped in a harmonic trap of transverse frequencies
$\omega_{ri}$, following the calculations in \cite{ho2009opdtq}.
Using Gibbs-Duhem relation at a constant temperature $T$,
$\textrm{d}P=\sum_in_i \textrm{d}\mu_i$, then
$$
\sum_i\frac{m_i\omega_{ri}^2}{2\pi}\overline{n}_i=\int\sum_i
\frac{m_i\omega_{ri}^2}{2\pi}\textrm{d}x\textrm{d}y \frac{\partial
P}{\partial\mu_i}=\int \sum_i \textrm{d}\mu_i\frac{\partial
P}{\partial \mu_i},
$$
where we have used local density approximation ($\mu_i(\mathbf{r})=\mu_i^0-V(\mathbf{r})$) to
convert the integral over space to an integral on the chemical
potentials. The integral is straightforward and yields to
$$
P(\mu_{iz},T)=\frac{1}{2\pi}\sum_im_i\omega_{ri}^2\overline{n}_i(z).
$$\\

\end{document}